\documentclass[a4paper,11pt]{article}

\usepackage{jinstpub} 

\usepackage{lineno}
\usepackage{siunitx}
\usepackage{booktabs} 

\title{\boldmath A facility for radiation hardness studies based on a medical cyclotron}

\author{John Anders,}
\author{Saverio Braccini,} 
\author{Tommaso Stefano Carzaniga,}
\author{Pierluigi Casolaro,}
\author{Meghranjana Chatterjee,}
\author{Gaia Dellepiane,}
\author{Laura Franconi,}
\author{Lea Halser,}
\author[1]{Armin Ilg,\note{now at University of Zurich.}}
\author[2]{Isidre Mateu,\note{Corresponding author.}}
\author[3] {Federico Meloni, \note{now at Deutsches Elektronen-Synchrotron DESY, Hamburg.}}
\author[4] {Claudia Merlassino, \note{now at University of Oxford.}}
\author{Antonio Miucci,}
\author{Roman M\"uller,}
\author[3] {Marco Rimoldi,}
\author{and Michele Weber}

\affiliation{Albert Einstein Center for Fundamental Physics (AEC), Laboratory for High Energy Physics (LHEP),\\Sidlerstrasse~5, CH-3012 Bern, Switzerland}
\emailAdd{isidro.mateu@lhep.unibe.ch}

\abstract{The development of instrumentation for operation in high-radiation environments represents a challenge in various research fields, particularly in particle physics experiments and space missions, and drives an ever-increasing demand for irradiation facilities dedicated to radiation hardness studies. Depending on the application, different needs arise in terms  of particle type, energy and dose rate. In this article, we present a versatile installation based on a medical cyclotron located at the Bern University Hospital (Inselspital), which is used as a controlled 18-\si{\MeV} proton source. This accelerator is used for daily production of medical radioisotopes, as well as for multidisciplinary research, thanks to a 6.5-meter long beam transfer line that terminates in an independent bunker, dedicated only to scientific activities. The facility offers a wide range of proton fluxes, due to an adjustable beam current from approximately \SI{10}{\pico \ampere} to the micro-ampere range, together with a series of steering and focusing magnets along the beamline that allow for the beam spot to be focused down to a few \si{\milli \meter ^{2}}. The beamline can be instrumented with a variety of beam monitoring detectors, collimators, and beam current measurement devices to precisely control the irradiation conditions.  The facility also hosts a well equipped laboratory dedicated to the characterisation of samples after irradiation. An experimental validation of the irradiation setup, with proton fluxes ranging from \SI{5e9}{\cm^{-2}\s^{-1}} to \SI{4e11}{\cm^{-2}\s^{-1}}, is reported.}


\begin{document}
\maketitle
\flushbottom
\section{Introduction}
Radiation Hardness Assurance (RHA) is a key aspect of scientific experiments carried out in high-radiation environments, where the replacement of defective equipment is not realistic from the point of view of cost, time, or access to the experimental area. Clear examples of such circumstances are space missions, in which systems must be designed to withstand the radiation levels expected during the whole mission's lifetime. To this end, ESA, NASA, and other space agencies carry out extensive R\&D efforts for RHA and define strict protocols for the selection and testing of materials and components for their missions \cite{furano2013review} \cite{poivey2002radiation} \cite{label1998emerging}. Similarly, the detector and accelerator components at the CERN Large Hadron Collider (LHC) \cite{Bruning:782076} must operate at extraordinary levels of radiation for several years before being replaced according to a detailed  plan of maintenance and upgrades. Specifically, the High-Luminosity upgrade of the LHC \cite{Aberle:2749422}, with the aim of increasing the integrated luminosity by a factor of five by 2030, poses a challenge in terms of RHA for the sensor technologies and electronics \cite{8116686} \cite{Moll:2018fol} \cite{DIMITRIEVSKA2020162091}. Medical  \cite{6551318} \cite{https://doi.org/10.1002/pssa.201800383} and nuclear \cite{VUKOLOV2015177} \cite{9217477} applications are examples of other fields where RHA is of  concern. 

\textcolor{black}{Three different mechanisms are responsible for radiation-induced damage to sensors and electronics: changes in digital electronics' logic states due to the interaction of a single ionising particle, commonly referred as single event effects, ionisation and trapping of charge in dielectrics such as the gate oxides present in integrated circuits, and dislocation of lattice atoms in semiconductors, also known as displacement damage (DD). While single event effects can occur at any time with a certain probability that depends on the particle flux, type and energy, charge trapping and DD are cumulative effects, i.e. their severity depends on the cumulated radiation exposure of the device. Two quantities are used to measure them: Total Ionising Dose (TID) and 1-MeV neutron equivalent particle fluence\footnote{Particle fluence, or fluence, is used as a synonim of cumulated (integrated) particle flux in this paper.} (\si{n_{eq}\per\cm^{2}}). TID quantifies the amount of energy transferred to the material in the form of ionisation, eventually leading to charge trapping in insulating interfaces, and is expressed as energy per unit of mass. The SI unit is the Gray (Gy), where 1 Gy = 1 J/kg. The 1-MeV neutron equivalent particle fluence quantifies DD, and is interpreted as the fluence of 1-MeV neutrons that would result in the same amount of DD as the real fluence delivered by an irradiation source of a given particle type and energy. The conversion factor (commonly known as hardness factor) to 1-MeV neutron equivalent fluence is given by the particle's Non-Ionising Energy Loss (NIEL) cross section. NIEL is defined as the amount of energy transferred to the material which does not result in ionisation, thus available for DD. Calculated values of NIEL cross sections in silicon for various particles can be found in \cite{Moll:2018fol}.}

\begin{figure}
	\centering
	\includegraphics[width=\textwidth]{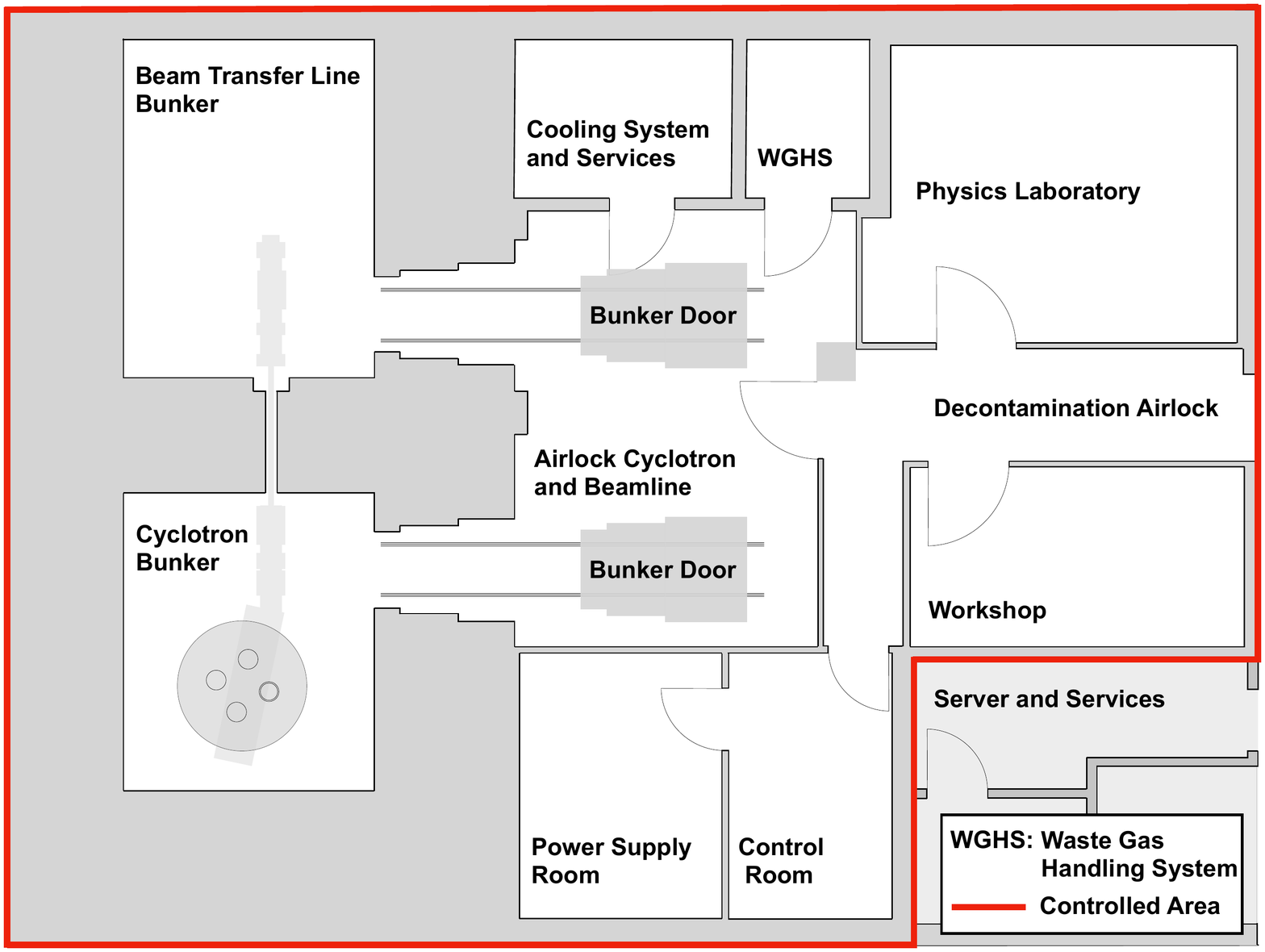}	
	\caption{ Layout of the cyclotron facility at the Bern Inselspital. It features two separate bunkers (depicted with their doors open in the figure) hosting the cyclotron and the BTL. In addition, a multi-purpose laboratory (Physics Laboratory) is available for post-irradiation studies.}
	\label{fig:irr-layout}
\end{figure}

The availability of dedicated irradiation facilities capable of delivering  controlled fluence and TID levels plays a crucial role in qualifying the radiation hardness of components. Such facilities are rare and in high demand, so access to them is often limited. Furthermore, the characterisation and analysis of samples after irradiation is normally carried out in specialised laboratories, away from the irradiation facility. This often requires the shipment of radioactive materials, which results in delays, additional costs and complications for the user. We present here a medical cyclotron facility located at the Bern University Hospital (Inselspital), featuring an 18-\si{\MeV} proton beamline dedicated to RHA tests and multidisciplinary research activities. A  convenient feature of this facility is the availability of a dedicated characterisation laboratory (Physics Laboratory) within the  radiation controlled area, which allows for post-irradiation studies on the device under test (DUT) to be performed directly in place and shortly after the irradiation. \textcolor{black}{In this way, not only the mentioned difficulties associated to the shipment of radioactive samples can be avoided, but also the scientific outcome can be improved by minimising any undesired thermal annealing on the irradiated samples}. A layout of the full facility is presented in Figure \ref{fig:irr-layout}. \textcolor{black}{RHA tests at the Bern medical cyclotron focus primarily on TID and DD effects, while the possibilities for SEE testing have not been investigated so far. 18-\si{\MeV} protons release a high amount of energy in matter in the form of ionisation (e.g. the electronic stopping power in silicon is $\sim$\SI{22}{\MeV \per (\g \per \cm^2)}\footnote{Electronic stopping power calculated using SRIM\cite{SRIM}.} compared to $\sim$\SI{2}{\MeV \per (\g \per \cm^2)} for a minimum ionising particle (MIP)), thus enabling radiation hardness tests at very high TID. On the other hand, the hardness factor for conversion to 1-MeV neutron equivalent fluence is approximately 1.4 \cite{Moll:2018fol}.\textcolor{black}{Practical information for users interested in  conducting irradiation experiments in the facility, as well as contact details, can be found on the University of Bern's Laboratory for High Energy Physics (LHEP) website\cite{cyclotron_website}.}}

\section{The Bern medical cyclotron and beam transfer line}

The Bern cyclotron is a commercial IBA Cyclone 18/18, accelerating H\textsuperscript{-} ions to a kinetic energy of \SI{18}{\MeV}   with a maximum extracted current of \SI{150}{\micro \ampere} \cite{Cyclo} \cite{Scampoli:2011zz}. Proton extraction is achieved by stripping the electrons from the accelerated H\textsuperscript{-} ions by means of thin pyrolytic carbon foils. The cyclotron is used routinely overnight to produce radioisotopes (namely $^{18}$F) for PET diagnostics, and is thus free for other research activities during the day. The on-going research programs include the production and study of novel radioisotopes for medical applications \cite{molecules25204706} \cite{braccini:cyclotrons2019-tua02}, the development of beam-monitoring devices \cite{Belver-Aguilar:2020plz} \cite{app10228217}, radiation protection studies \cite{PMID:24982259} and radiation hardness studies, which are the subject of this article.

\begin{figure}
	\centering
		 \includegraphics[width=0.8\textwidth]{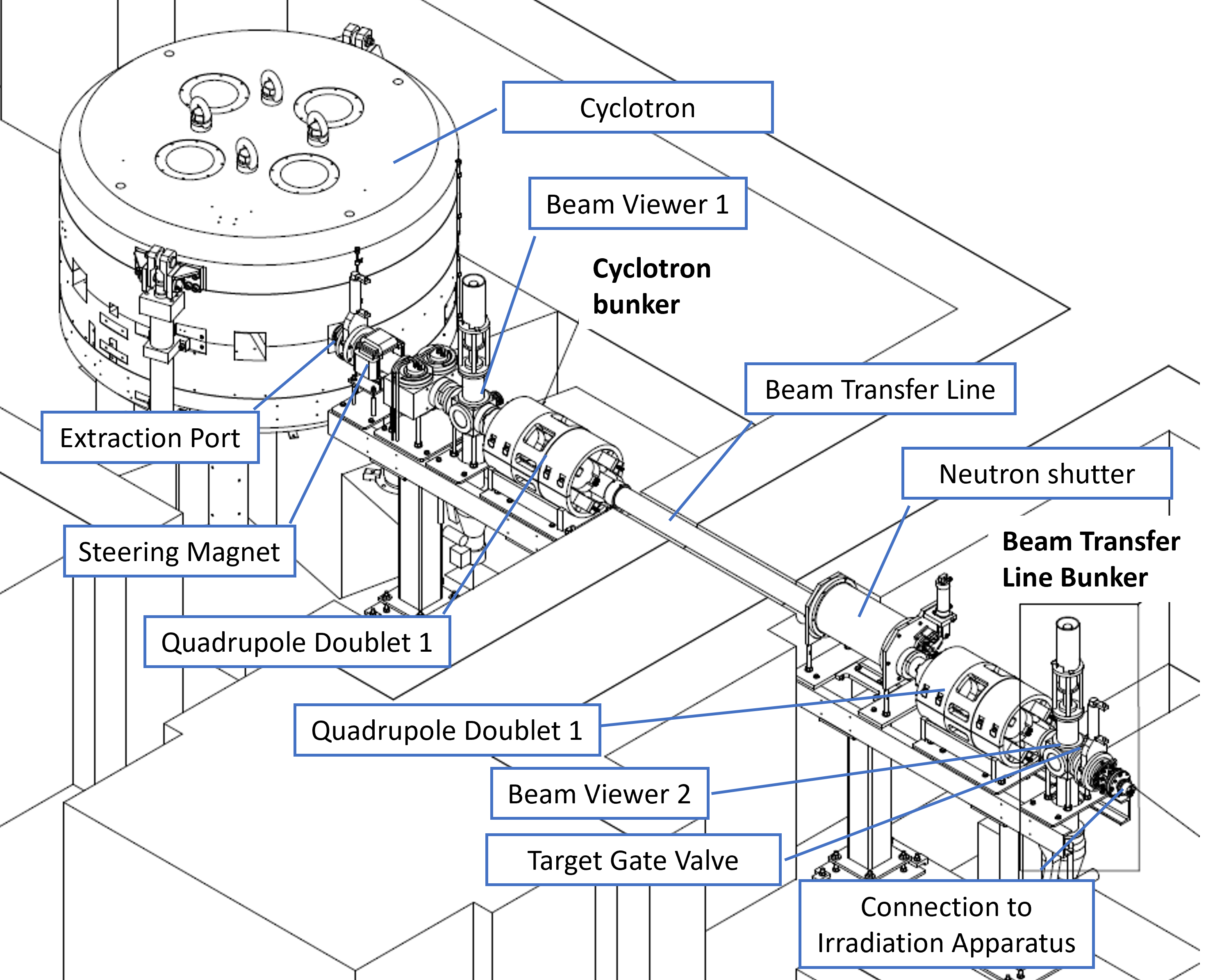}
		 \caption{Scheme of the cyclotron and irradiation bunkers. The cyclotron bunker is separated by a 1.8-\si{\meter} thick concrete wall from the BTL bunker where the irradiation apparatus is located. This allows for easy access to the DUT shortly after irradiation, due to the low level of activity in the BTL bunker (below \SI{2}{\micro \sievert \per \hour}).}
		 \label{fig:Cyclo_scheme}
\end{figure}

To perform radiation hardness tests, the extracted beam is transported via a 6.5-\si{\meter} long beam transfer line (BTL) \cite{Cyclo} into a separate bunker, referred to as the BTL bunker in this document, as depicted in Figure \ref{fig:Cyclo_scheme}.   The accelerated protons encounter first an X-Y steering magnet system and are afterwards focused using two,  horizontal and vertical, quadrupole doublets: the first located in the cyclotron bunker and the second in the BTL bunker. Two movable, water-cooled, beam viewers are present along the beam line, providing a destructive measurement of the beam current with a sensitivity of \SI{100}{\nano \ampere}. A transmission efficiency  greater than $95\%$ is typically achieved from the beam current measured on the stripper to the one measured on the second beam viewer. The BTL is terminated with a target gate valve to preserve the vacuum in the beam pipe during the mounting of scientific equipment.

A 1.8-\si{\meter} thick concrete wall separates  the two bunkers and prevents the  neutron flux generated during the  production of medical radioisotopes from reaching the BTL bunker. Furthermore, a  cylindrical neutron shutter is inserted in the beam pipe while the BTL is not in use. Thanks to this two-bunker structure, the dose rate in the BTL bunker during the  production is kept below \SI{2}{\micro \sievert \per \hour}.

To use the cyclotron for irradiation purposes, a dedicated beam characterisation was performed. In particular, the beam energy distribution was studied with several methods: using a multi-leaf Faraday cup \cite{instruments3010004}, measuring the beam transmission through aluminum absorbers of different thicknesses \cite{BernEnergy} and measuring the deflection angle introduced by a dipole electromagnet \cite{Haffner:2706007}. Additionally, the cyclotron operation settings were tuned for low output current operation, down to the \si{\pico\ampere} range\cite{Cyclo_LC}. \textcolor{black}{The need for such a low beam current was originally driven by radiobiology applications, while beam currents of the order of hundreds of nanoamperes are more typical during radiation hardness tests. However, being able to deliver beams over a wide range of intensities is of interest for applications where not only the total dose, but also dose rate effects, are of concern.} It is worth noting that medical cyclotrons for radioisotope production are conceived to operate at currents larger than \SI{10}{\micro \ampere}, and that the standard equipment (such as the beam viewers) is designed for this intensity range. In order to precisely determine the fluence and dose delivered to the DUT, it is fundamental to  measure the beam current and monitor the beam profile during the irradiation. In order to do so, several dedicated instruments, presented in the next section, were developed at LHEP. The entire irradiation setup has been validated for a proton flux ranging from \SI{5e9}{\cm^{-2}\s^{-1}} to \SI{4e11}{\cm^{-2}\s^{-1}}.

\section{The irradiation facility}

The BTL ends in the BTL bunker, a $5.4 \times 4.0$ \si{\meter^{2}} room dedicated to  research activities. Approximately 3 meters from the end of the BTL to the bunker wall are available for the irradiation setup. An irradiation table, an X-Y motorised stage system, as well as multiple tripodes, supports, and lead bricks are available for the installation of the DUT. Furthermore, the room is equipped with \SI{220}{\volt} power sockets, Ethernet, fibre optic, BNC and SHV connections to the Physics Laboratory, enabling the active readout of the device during irradiation. Temperature control on the DUT is also possible with Peltier elements and/or water cooling.

\begin{figure}
	\centering
\includegraphics[width=0.9\textwidth]{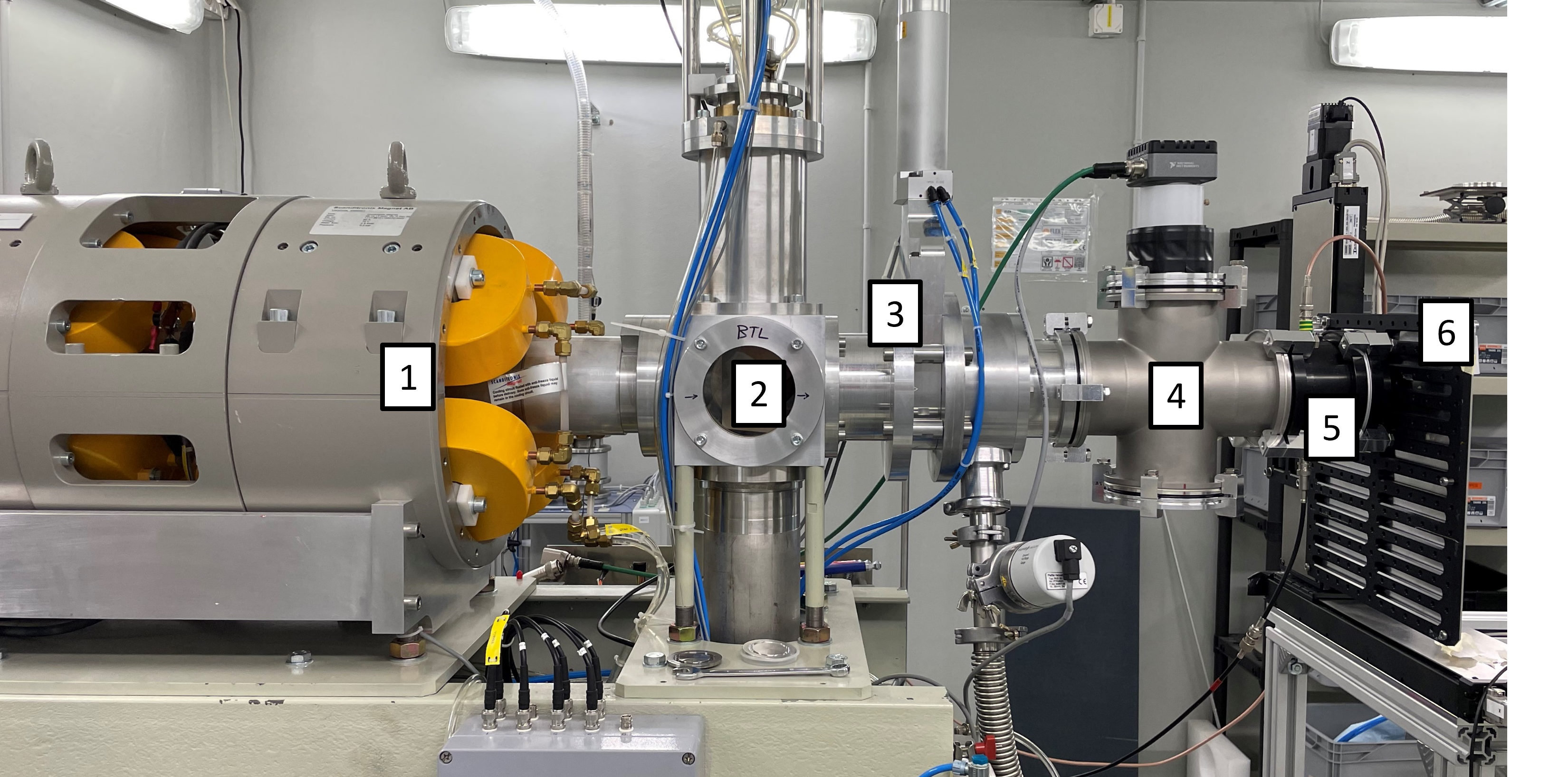}
	\caption{Irradiation setup. It consists of a quadrupole doublet~(1), a beam viewer~(2), the target vacuum valve~(3), the two-dimensional beam profile monitor $\pi^2$~(4) and the collimator terminated with an aluminium extraction window (5). The DUT is mounted and aligned on the remote controllable 2D stage (6).}
	\label{fig:irr-setup}
\end{figure}

Irradiations are typically carried out in air (although irradiations in vacuum are also possible) using the configuration shown in Figure \ref{fig:irr-setup}. Right after the vacuum valve that terminates the BTL, a beam profile monitor is installed to measure the transverse beam profile, followed by a collimator  that also acts as a beam current measuring device. The collimator aperture can be changed, depending on the desired beam spot size, with a maximum of $30 \times 30$ \si{\milli \meter^{2}}. Real-time monitoring of the proton flux delivered to the DUT is achieved by combining the beam intensity and profile measurements. \textcolor{black}{After the collimator, the beam is extracted in air through a 300-\si{\micro \meter} thick aluminium window. The DUT is typically placed as close to the extraction window as possible, in order to minimise the distance travelled in air by the protons. As a radioprotection measure, the radioactivity in air is continuously monitored in the facility \cite{PMID:24982259}.}

\textcolor{black}{The DUT can be installed on the remotely controlled two-dimensional motorized stage system shown in Figure \ref{fig:irr-setup}. Multiple DUTs can be mounted on the stage system and irradiated consecutively without need to enter the BTL bunker. Stacking of multiple DUTs is rarely possible due to the beam's short radiation length in most materials. The stage system features a $30 \times 30$ cm$^2$ grid of M6 threaded holes spaced with a \SI{2.5}{\cm} pitch, and can hold up to \SI{8}{\kg}. Larger samples can be installed on other supports such as tripods or tables, also available in the facility. A laser system for the horizontal and vertical alignment of the DUT with the centre of the beam pipe is installed in the bunker.}


\subsection{Collimation and beam intensity measurement}
\label{sec:collimator}

\begin{figure}
	\centering
\includegraphics[width=0.95\textwidth]{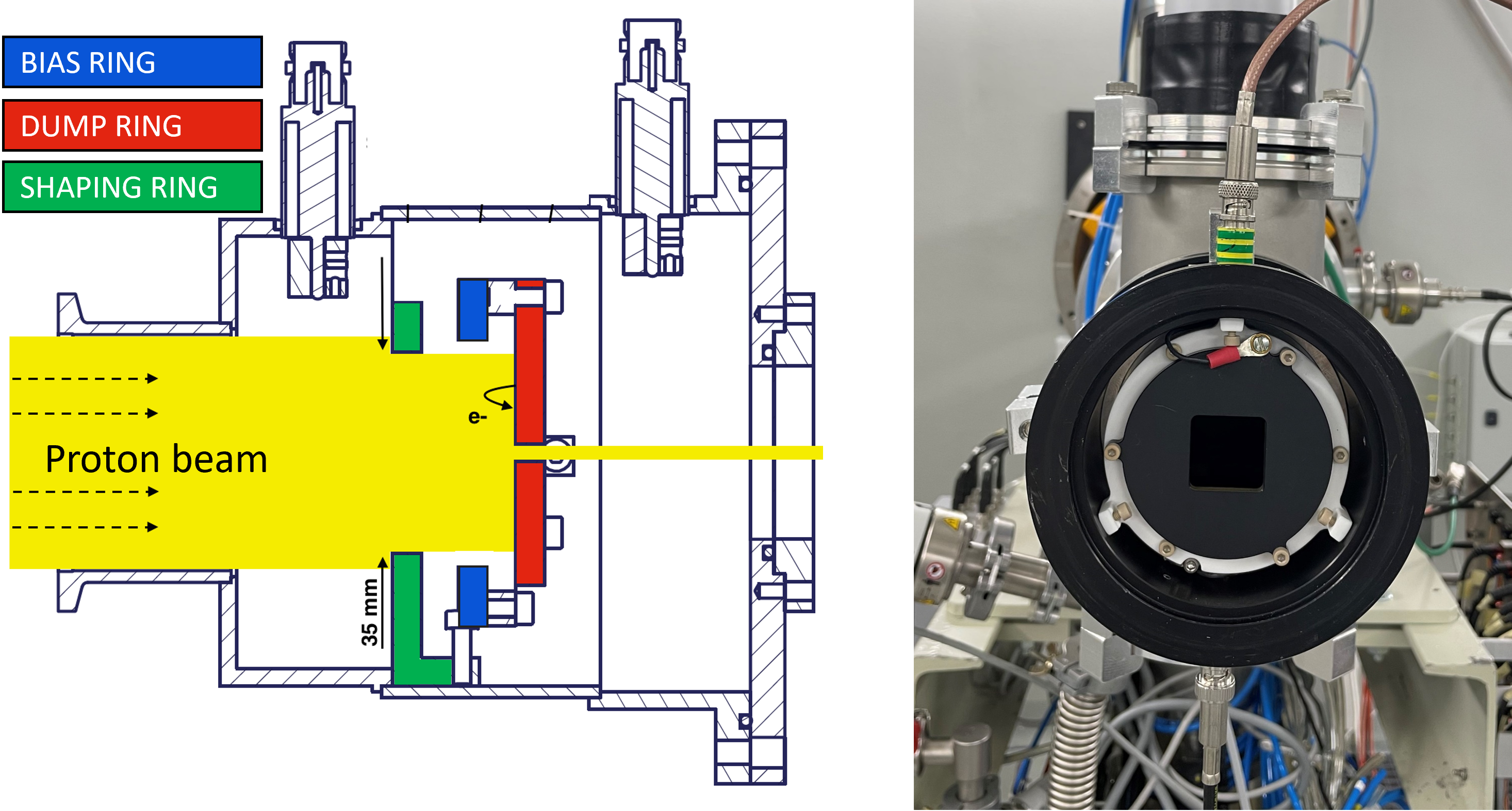}
  \caption{Cross-section of the current measuring collimator (left) and view of the instrument from the point of view of the DUT (right).} 
  \label{fig:collimator}
\end{figure}

To measure the beam current, a dedicated collimator was developed. Its cross-section is shown in Figure \ref{fig:collimator}.  It consists of three aluminium rings: the shaping ring, the bias ring, and the dump ring, each of them 5-\si{\milli\meter} thick (18-\si{\MeV} protons have a range of \SI{1.6}{\milli\meter} in aluminium). The proton beam enters the collimator through a 40-\si{\milli\meter} diameter aperture and is first shaped by the shaping ring  (35-\si{\milli\meter} diameter). Then, the dump ring stops the outer part of the beam, while the central part is collimated according to the chosen aperture (several versions of the dump ring are available, with aperture sizes from $5 \times 5$ \si{\milli \meter^{2}} to $30 \times 30$ \si{\milli \meter^{2}}). The dump ring is electrically connected to an electrometer in order to measure the intercepted beam. A \SI{100}{\volt} negative potential, applied to the biasing ring, prevents Secondary Electron Emission (SEE) from the dump ring, which would lead to an up to 20\% current overestimation, as documented in \cite{Cyclo_LC}. The collimator is connected at the end of the BTL and is operated in vacuum to prevent electrons from ionised air molecules from being collected by the dump ring.

\subsection{Beam profile monitoring}

Two different devices, the so called UniBEaM \cite{UniBeam} and $\pi^2$ \cite{Belver-Aguilar:2020plz}  detectors, were developed for monitoring the beam profile and position. Both are enclosed in vacuum-tight structures and can be easily installed in the beam line according to specific needs. Typically, the beam profile monitor is installed upstream from the collimator described in the previous section, in order to measure the beam footprint over the whole collimator's dump ring area.

\begin{figure}
	\centering
	\includegraphics[width=\textwidth]{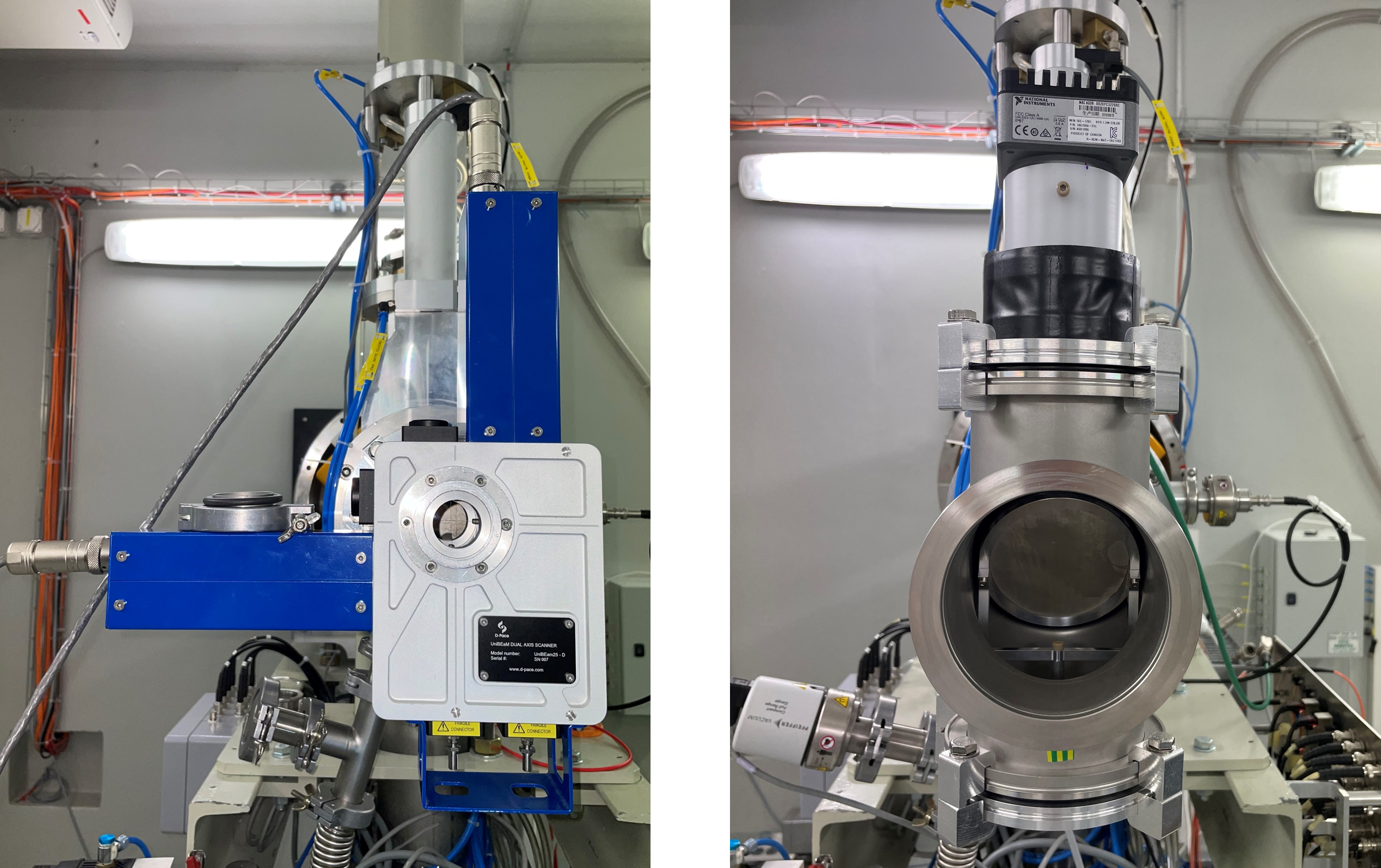}
	\caption{The commercial version of the UniBEaM (left) \cite{UniBeam}  and the $\pi^2$ (right) \cite{Belver-Aguilar:2020plz} beam profile monitors at the Bern cyclotron. Both devices are seen from the point of view of the DUT.}
	\label{fig:irr-bpm}
\end{figure}

\begin{figure}
	\centering
	{\includegraphics[width=1.0\textwidth]{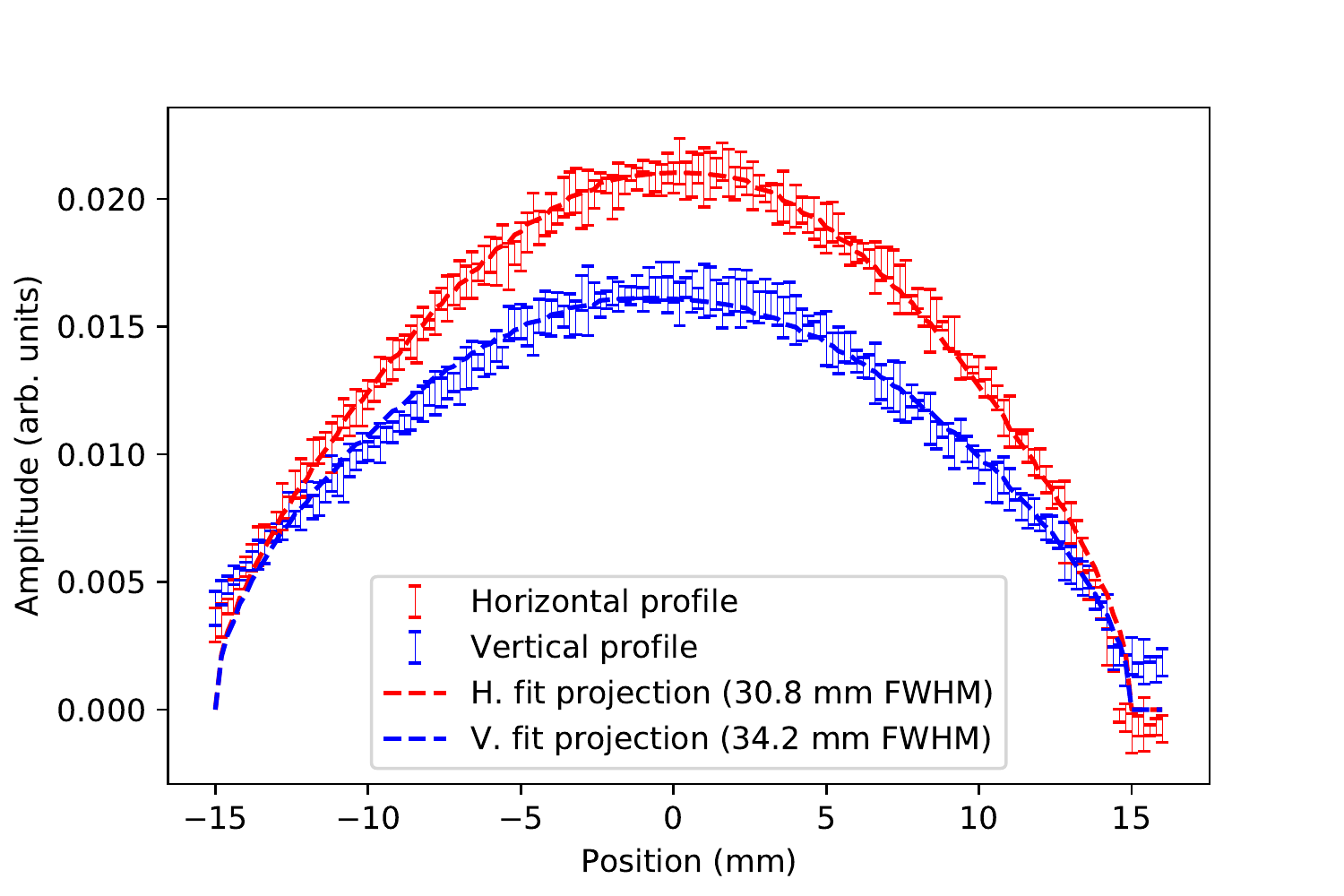}}
  	\caption{Beam profile at the end of the BTL, measured by the UniBEaM profile monitor.}
	\label{fig:unibeam_profiles}\quad
\end{figure}

The UniBEaM (Figure \ref{fig:irr-bpm}, left) uses doped-silica optical fibres (200-\si{\micro \meter} diameter) as scintillating elements. In order to determine the horizontal projection of the transverse profile, one such fibre, vertically oriented, is continuously moved along the horizontal axis by means of a stepper motor. The range of movement covers the full 3-\si{\centi\meter} diameter circular aperture of the instrument. Part of the scintillating light propagates in the fibre and is transmitted via a fibre optic cable to a photo-detector located outside the BTL bunker. Dedicated software controls the motor movement and retrieves the light intensity information to reconstruct and display the beam profile in real time. The vertical profile is measured in an analogous fashion by a second fibre, oriented horizontally and moved along the vertical axis. The motor speed is such that it takes only a few seconds for a fibre to scan its full range of motion. The fibre movement can be turned off after the initial tuning of the beam, to avoid any disturbance on the beam during irradiation. \textcolor{black}{Figure \ref{fig:unibeam_profiles} shows an example of the typical profile achieved with the BTL magnets configured for broad beam conditions. Under these conditions, the beam footprint is well described by a two-dimensional Gaussian probability density function (PDF). The horizontal and vertical projections of a fit to this PDF are also included in the figure}. The laboratory is equipped with several UniBEaM detectors produced by LHEP and a commercial version by the canadian company D-PACE (under licence from the University of Bern) \cite{DPACE}. 

\begin{figure}
	\centering
	{\includegraphics[width=1.0\textwidth]{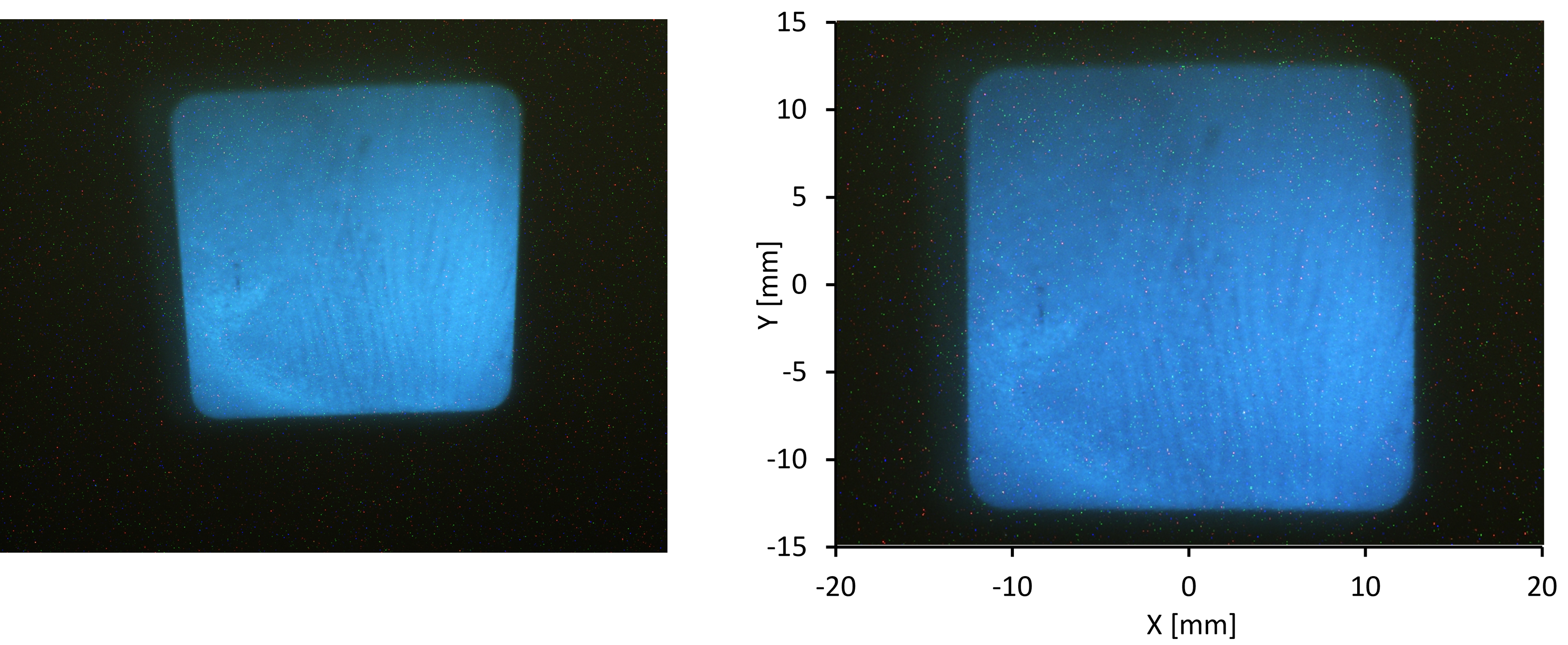}}
  	\caption{Left: raw image of the beam footprint obtained with the $\pi^2$ beam profiler. Right: same image, after post-processing to correct for perspective effects.}
	\label{fig:2D_profile}\quad
\end{figure}

The $\pi^2$ detector  (Figure \ref{fig:irr-bpm}, right)  consists of a scintillating foil installed at the centre of a 4-way vacuum cross. The foil is a 15-\si{\micro \meter} thick aluminium, coated with a cerium and terbium doped yttrium orthosilicate (Y$_2$SiO$_5$:Ce,Tb) Phosphor screen (P47 compound), and is placed at a \ang{45} angle with respect to the beam axis. The beam footprint on the scintillator is imaged with an ISC-1781 camera from National Instruments, looking at the foil through an optical window installed on the upper port of the cross.  The foil covers a 6-\si{\cm} diameter circular area orthogonal to the beam axis, thus enabling the observation of the full beam profile in any configuration. An earlier version of the same detector, covering only a 4-\si{cm} diameter \cite{Belver-Aguilar:2020plz}, is also available. An Ethernet connection from the camera to a PC in the control room allows the image to be retrieved and processed using commercial software from National Instruments. The software allows to apply a transformation matrix to the image in order to correct for the perspective effect due to the \ang{45} orientation of the camera with respect to the foil. A dedicated calibration  using millimeter paper instead of the scintillating foil was performed to obtain the appropriate transformation matrix. \textcolor{black}{Figure \ref{fig:2D_profile}, demonstrates the perspective effect in the raw image captured by the camera (left) and the applied correction (right). To obtain this image, the normal order of the beam profile monitor and collimator was inverted: the collimator, with a $2.5 \times 2.5$ cm$^2$ aperture, was installed upstream from the $\pi^2$.}

\textcolor{black}{Note that, because of its larger aperture, the  $\pi^2$ allows to monitor broader beams than the UniBEaM. (6 cm versus 4 cm diameter). Moreover, a greater sensitivity can be achieved with the $\pi^2$ by increasing the camera exposure time. One drawback of this detector, however, is the potential non-uniform light yield of the scintillating foil either due to an irregular initial deposition of P47, or to a degradation after long term irradiation exposure. Corrections to this effect can be achieved by performing periodical calibrations against other beam profile measurements (e.g. the UniBEaM detector or radiochromic films).} 

\subsection{Beam delivery and proton flux measurement}
\label{sec:fluence_assesment}
\begin{figure}
\centering
		\includegraphics[width=0.8\textwidth]{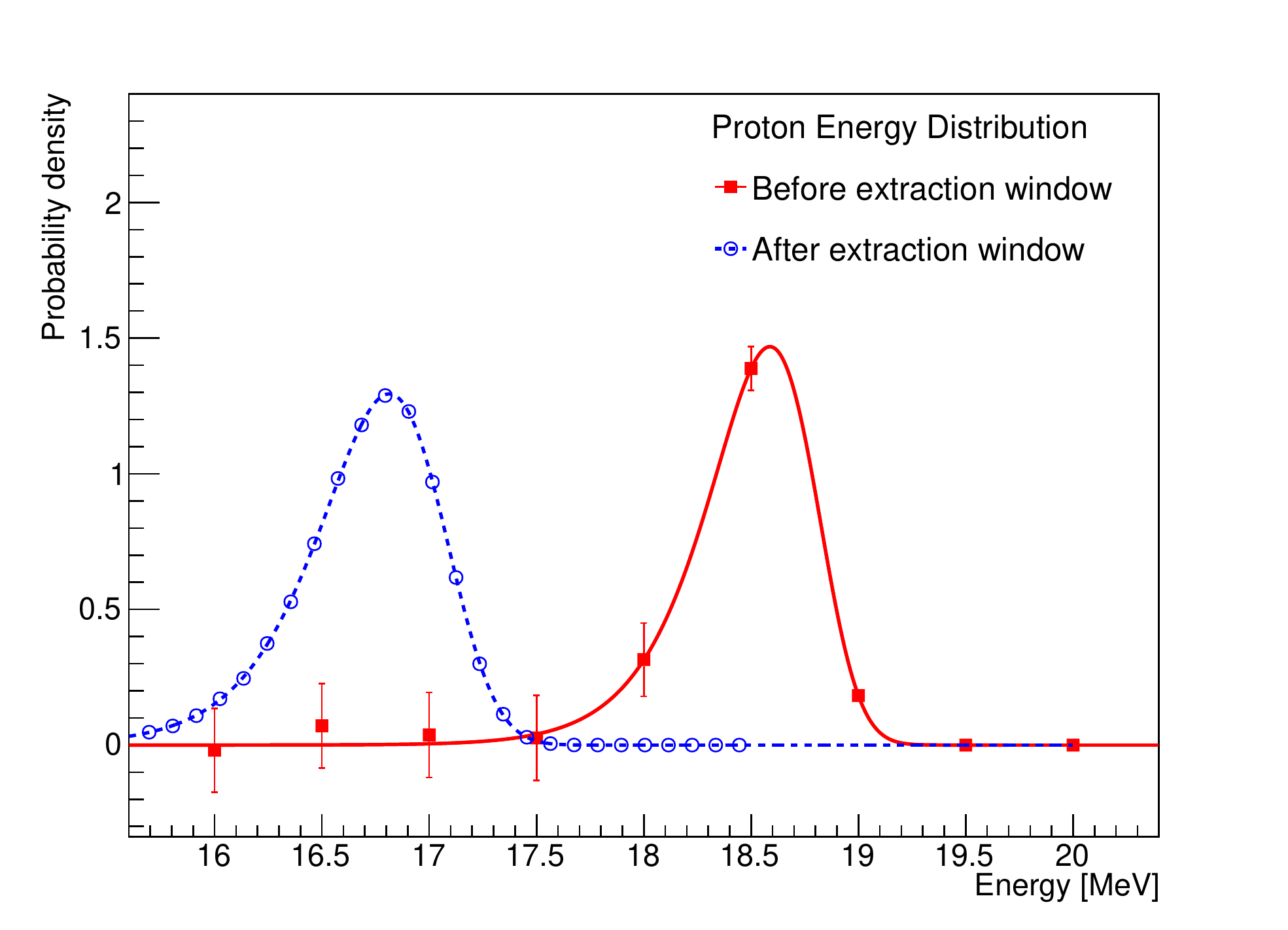}
		\caption{  Distribution of the proton energy before and after extraction. The square points are the measured energy profile of the beam before extraction \cite{BernEnergy}, fitted with a Verhulst function. The effect of a  $300\,\si{\micro\meter}$ aluminium extraction window has been simulated and is shown by the circle markers.}
		\label{fig:BeamEnergy}
\end{figure}

Downstream from the beam profile monitor and the collimator, the beam is extracted from vacuum to air and onto the DUT through a 300-\si{\micro \meter} thick aluminium  window. To study the effect on the beam energy, the energy PDF obtained in a previous measurement \cite{BernEnergy}, and shown by the square points in Figure\,\ref{fig:BeamEnergy}, was considered. The PDF was fitted with a Verlhust function shown by the red curve. The stopping power of aluminium for protons, calculated with SRIM  \cite{SRIM}, was used to derive the energy PDF after the extraction window, represented by the blue circles in Figure\,\ref{fig:BeamEnergy}. A fit to a Verhulst function convoluted with a Gaussian resulted in a mean energy of $16.7\pm0.5\,$\si{\MeV}. For reference, Table \ref{table:parameters} shows the electronic and nuclear stopping power of different materials for a $16.7\,$\si{\MeV} proton beam.

\begin{table}
\caption{Electronic stopping power for different materials in case of a \SI{16.7}{\MeV} proton beam. The values are obtained from the NIST database \cite{NISTdatabase}.}
\label{table:parameters}
\centering
\begin{tabular}{l|cc}
\toprule
Material  & $\dfrac{1}{\rho}\dfrac{dE}{dx}_{\rm{ion}}\,\big[\si{\MeV} \frac{\si{\centi\meter}^2}{\si{\gram}}\big]$ & $\dfrac{1}{\rho}\dfrac{dE}{ dx}_{\rm{nuc}}\,\big[ \si{\MeV}\frac{\si{\centi\meter}^2}{\si{\gram}}\big]$\\
\midrule
Silicon & 23.25 & 0.011\\
Silicon dioxide & 24.31 & 0.012\\
Aluminium & 22.67 & 0.011 \\
Plastic scintillator (Vyniltoluene) & 30.12 & 0.015 \\
Mylar & 28.06 & 0.013 \\
Photographic emulsion & 18.30 & 0.009 \\
\bottomrule
\end{tabular}
\end{table}

The measurement of the beam current during irradiation allows for the monitoring of the proton flux in real time,  and is automatically recorded for offline analysis. In particular, the dose rate measurement is relevant to identify non-linear effects in electronics, as discussed in \cite{FEI4_irr}. Assuming a uniform beam profile, the proton flux delivered to the DUT can be calculated from the beam current measured on the collimator as:

\begin{align}
\centering
\phi(t) &= \frac {j(t)}{Q_{\rm{e}}} \, \label{eq:fluence}
\end{align}

\noindent where $j(t)$ is the current per unit area, and $Q_e$ is the elementary charge. Assuming an incoming beam with a uniform flux density, $j(t)$ is calculated from the ratio between the current measured on the dump ring of the collimator system and its area. A close-to-uniform (within 20\%) profile is achieved by switching off the last quadrupole doublet in the BTL line.

A better estimation of the flux can be obtained by taking into account the measured beam profile. In this case, the flux can be calculated as:

\begin{align}
\centering
\phi (t) &= \frac{\int_{A_{\rm{DUT}}} P(x,y) dS}{\int_{A_{\rm{coll}}} P(x,y) dS}\cdot\frac{A_{\rm{coll}}}{A_{\rm{DUT}}} \cdot \frac {j(t)}{Q_{\rm{e}}}  \, \label{eq:factor}
\end{align}

\noindent where $P(x,y)$ is the transverse beam profile measured by the beam monitor, and $A_{\rm{DUT}}$ and $A_{\rm{coll}}$ are the areas of the DUT and the collimator (which varies according to the selected aperture), respectively.

A dedicated simulation was performed to estimate the uncertainty associated to this flux calculation method, considering  $P(x,y)$ to be the 2D Gaussian fit of the UniBEaM measured profile with typical broad beam settings, shown in figure \ref{fig:unibeam_profiles}. The simulation considered the statistical fluctuations on the current measurements, the uncertainties on the beam profile fit parameters,  \textcolor{black}{ and conservative estimates of \SI{500}{\micro \meter} (one sigma) for the tolerance on the dump ring diameter and aperture dimensions, and 1 mm (one sigma) for the tolerance on the alignment of the collimator, the UniBEaM and the DUT.} The overall uncertainty determined by the simulation was of approximately 10\%.

\begin{figure}[t]
\centering
		 \includegraphics[width=0.8\textwidth]{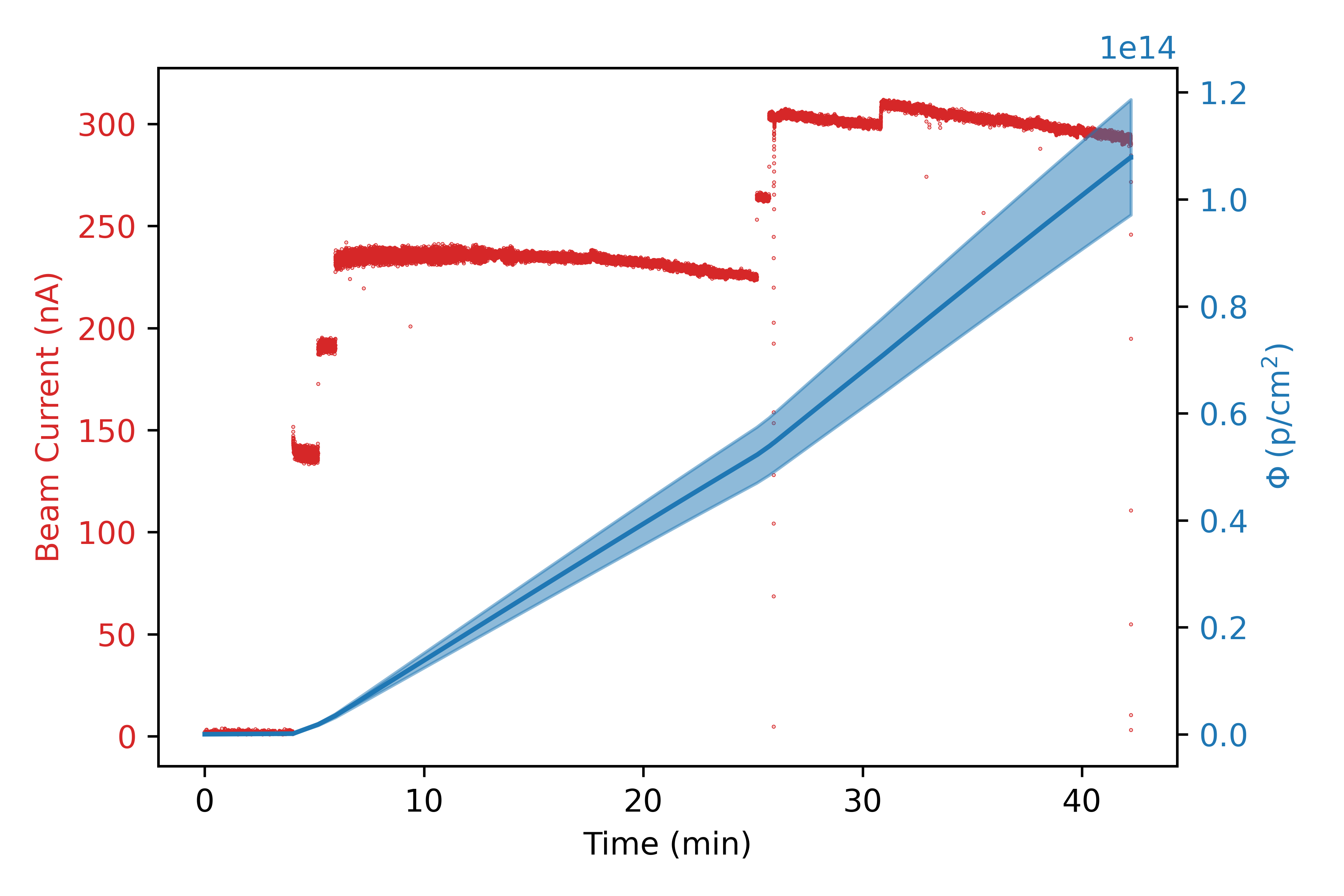}
  \caption{ Observed beam current during an irradiation (red trace), together with the  proton fluence delivered to the DUT (blue trace, the shaded area represents the $10 \%$ uncertainty on the fluence measurement). The steps in the beam current are due to the beam tuning performed by the cyclotron operator.}
\label{fig:current_plot}
\end{figure}

Figure~\ref{fig:current_plot} shows an example of the beam current intercepted by the dump ring and the proton fluence during a typical irradiation. The blue curve represents the fluence with the quoted 10\% uncertainty. \textcolor{black}{The beam intensity downward drift over time is due to the warm-up of the cyclotron main coil, which causes a variation of the magnetic field intensity and thus a loss of resonance with the electric field's alternating frequency. This effect is mitigated by the operator by periodically adjusting the main coil current. Ultimately, a beam current stability over time of approximately $10\%$ can be easily achieved.}

\subsection{Validation of the irradiation setup}

To validate the proton flux calculation method described in the previous section, two independent tests were carried out: the first using a Faraday cup, the second using calibrated radiochromic films.

\subsubsection{Validation using a Faraday Cup}
\begin{figure}
\centering
		 \includegraphics[width=0.8\textwidth]{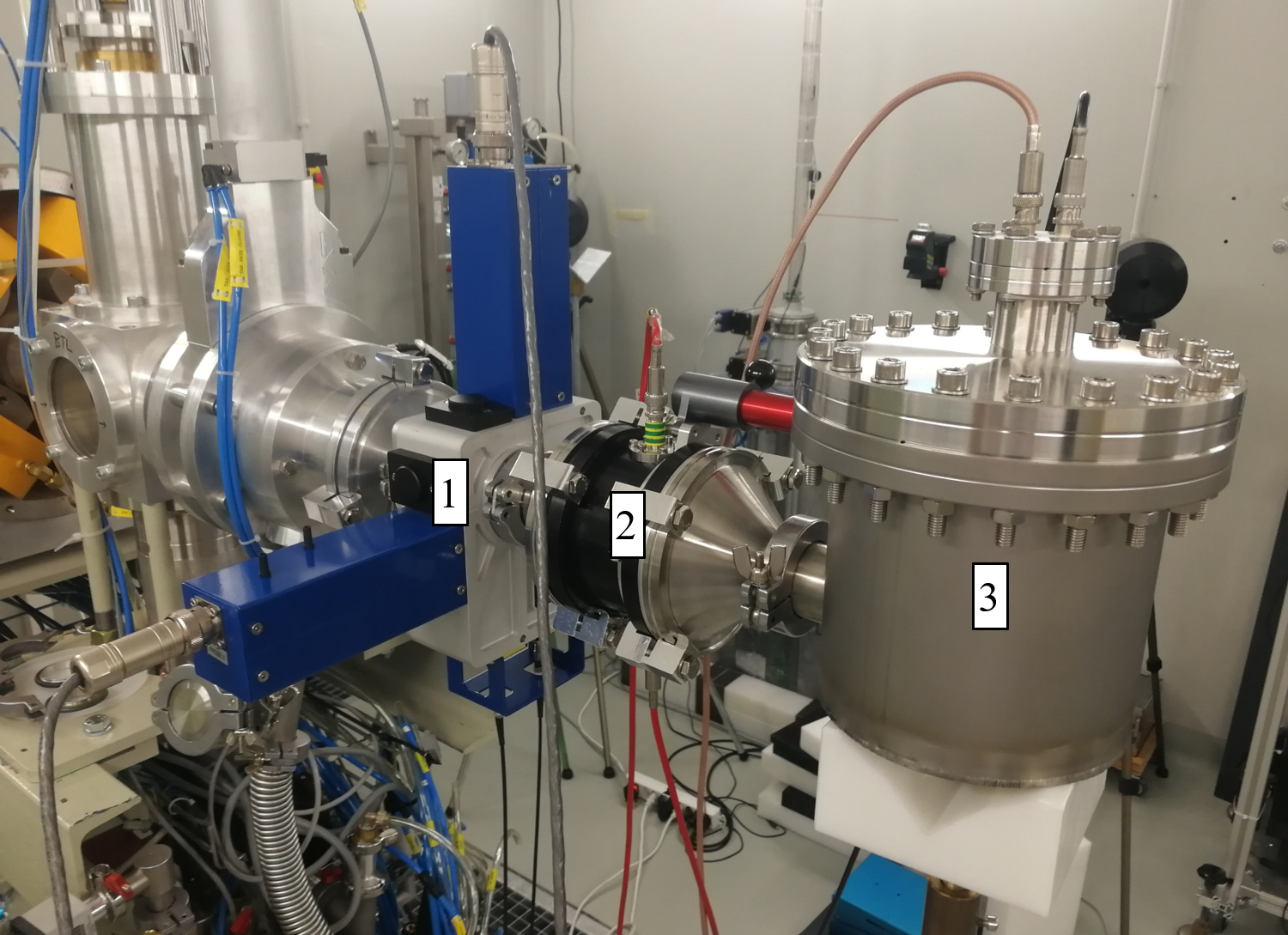}
  \caption{Validation of the irradiation setup. The UniBEaM  (1) was installed before the collimator (2), mimicking a typical irradiation setup. A Faraday cup (3) was installed at the DUT position to measure the beam current.}
\label{fig:coll_calibration}
\end{figure}

In this experiment, a high-sensitivity Faraday cup was installed after the collimator to obtain a direct measurement of the beam current at the DUT position. A picture of the setup is shown in Figure \ref{fig:coll_calibration}. The beam profile monitor was installed as first device at the end of the BTL (although only the UniBEaM is shown in the picture, the experiment was performed also with the $\pi^2$ device), immediately followed by the collimator and the Faraday cup. Following the same working principle of the collimator, described in Section \ref{sec:collimator}, the Faraday cup is also equipped with a repelling electrode to counteract SEE. In this experiment, a common \SI{100}{\volt} negative potential was applied to both repelling electrodes (collimator and Faraday cup). For both devices, the effect of the biasing voltage saturates above \SI{50}{\volt}. 

Equation \ref{eq:factor} was used to calculate the proton flux on the DUT. In the case of the UniBEaM, $P(x,y)$ was obtained from the 2D Gaussian fit on the measured horizontal and vertical beam profiles, while for the $\pi^2$, the 2D beam profile is directly available from the captured image.

\begin{figure}
\centering
		 \includegraphics[width=0.8\textwidth]{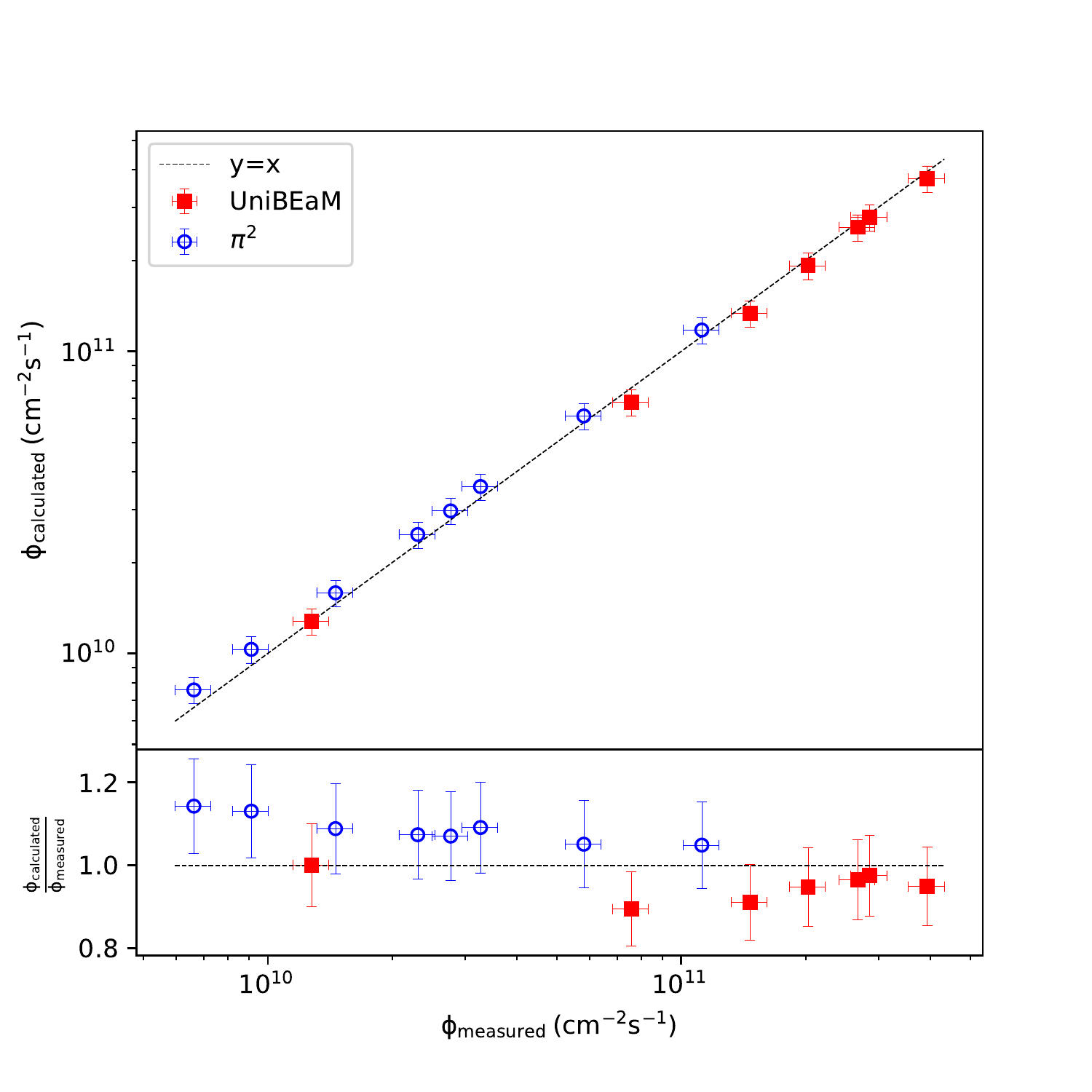}
  \caption{ Proton flux obtained using Equation \ref{eq:factor} for the UniBEaM (red squares) and the $\pi^2$ (blue circles), as a function of the flux obtained from the beam current measured on the Faraday cup. To guide the eye, the dashed line represents the identity function. For both beam profilers, the calculated proton flux is compatible within 10\% with the Faraday cup measurement, as shown by the ratio in the bottom plot.}
\label{fig:coll_calibration_results}
\end{figure}

Figure \ref{fig:coll_calibration_results} compares, for different primary beam intensities, the calculated proton flux versus the flux measured with the Faraday cup (obtained  by dividing the Faraday cup current by the collimator aperture). The results for both beam profile monitors show similar performance, compatible with the 10\% uncertainty on the calculated flux, as described in Section \ref{sec:fluence_assesment}. 

\subsubsection{Validation using radiochromic films}

In this experiment, a set of  calibrated radiochromic films were irradiated in incremental time intervals, with the goal of comparing the TID extracted from the change in colouration of the film, with the expected TID based on the calculated proton fluence, obtained by integrating the flux calculated using Eq. \ref{eq:factor}. FWT-60 radiochromic films from Far West Technology (FWT)~\cite{FWT} were used. They consist of  of hexa(hydroxyethyl) aminotriphenylacetonitrile (HHEVC) dyes, and are available in different thicknesses ($10\,\si{\micro\meter}$ and $47\,\si{\micro\meter}$).

Since the thickness of the film is much smaller than the proton range in the material, the energy loss of the protons in the film can be neglected and the TID ($\mathcal{D}_{\rm{ion}}$) can be calculated from the integrated proton flux:

\begin{align}
\centering
&\mathcal{D}_{\rm{ion}} =  \dfrac{1}{\rho} \dfrac{dE}{ dx}_{\rm{ion}}\cdot \int_{t_0}^{t_1}\phi(t) dt  \, \label{eq:TID_est} 
\end{align}

where $\dfrac{dE}{ dx}_{\rm{ion}}$ is the electronic stopping power of the material for protons, $\rho$ the material density, $\phi(t)$ the instantaneous proton flux and [t$_0$, t$_1$] the irradiation interval. From SRIM,  the stopping power of HHEVC  for a $16.7\,$\si{\MeV} proton beam is $30.87\,\si{\MeV}\cdot\si{\centi\meter}^2\si{\gram}^{-1}$.



\begin{figure}
\centering
\includegraphics[width=0.8\textwidth]{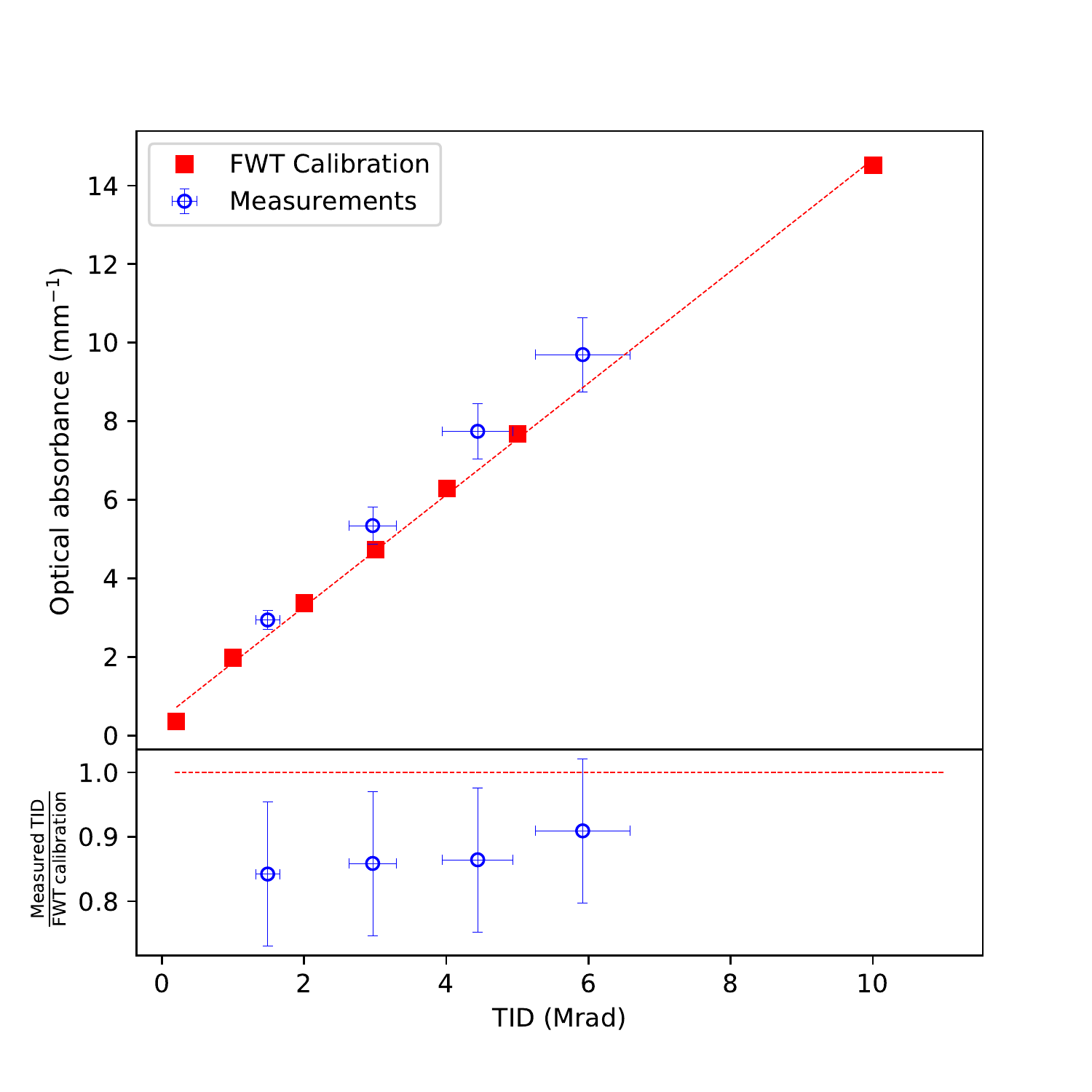}
\caption{Comparison between the TID measured using the method described in this document (blue circles) and the one determined from the change in colour of the radiochromic films, according to the calibration provided by FWT (red squares) \cite{FWT}. The two measurements are compatible within 15\% as shown by the ratio in the bottom plot.}
\label{fig:film_cal}
\end{figure}

 Figure\,\ref{fig:film_cal} shows the optical absorbance of the  films after irradiation in the BTL against the TID determined from Equations \ref{eq:factor} and \ref{eq:TID_est}
  (blue circles), compared with the calibration provided by the manufacturer (red squares). The bottom plot shows the ratio between the calculated dose and the one expected from the calibration at the same optical absorbance. To minimise the uncertainty on the optical absorbance several films were irradiated to each dose step in independent measurements. The uncertainty on the measured optical absorbance takes into account an uncertainty of $5\,\si{\micro\meter}$ on the film thickness, as indicated by FWT. \textcolor{black}{A 10\% uncertainty associated to the proton flux determination, as discussed for the previous measurement, was considered for the calculation of the TID. In addition, the uncertainty on the beam energy (16.7 $\pm$ 0.5 \si{\MeV}) translates into a $5\%$ uncertainty on the stopping power value. An agreement within $15\%$ is observed between the measurements and the calibration, thus validating the proton flux calculation method described in the previous section.}

\subsection{On site characterisation of the irradiated samples}

The Physics Laboratory is located within the Cyclotron radiation-controlled area. It is a space fully dedicated to the research activities conducted in the BTL bunker. From the Physics Laboratory the users can operate  their instruments in the BTL bunker via a patch panel featuring BNC, SHV, Ethernet, and different flavours of multi-pin and fibre optic connectors. NIM crates and electronics modules for signal acquisition, as well as common table-top electronics are also available in the laboratory. 

The Physics Laboratory also constitutes a space  to perform post-irradiation studies, often suppressing the need to transport radioactive material outside the facility. The laboratory equipment can be used for the characterisation of irradiated samples, but it is also possible for users to move their own characterisation setup to the facility. The laboratory is equipped with a HPGe gamma-spectrometer and an alpha-spectrometer, as well as a freezer for storage of irradiated samples at low temperature, in the case that undesired annealing effects need to be avoided.

\section{Examples of irradiation campaigns}

\begin{figure}
\centering
\includegraphics[width=\textwidth]{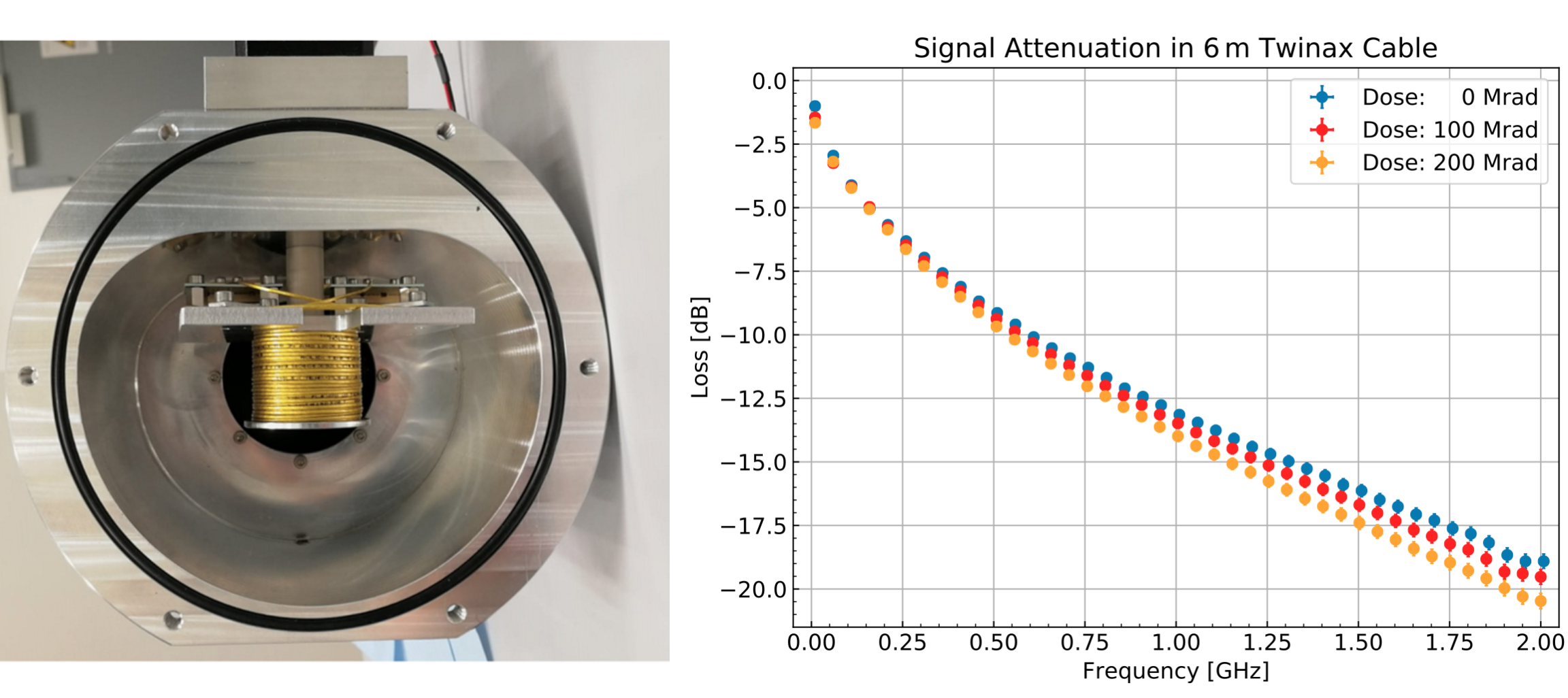}
\caption{Left: setup used to irradiate a 6-metre long twinax cable for the ATLAS ITk pixel readout chain; Right: Insertion Loss measured with a Vector Network Analyser at different dose levels.}
\label{fig:twinax}
\end{figure}

The Bern Medical Cyclotron is involved in the qualification of new sensor technologies, as well as other materials, in view of  the detector upgrades for the High-Luminosity LHC. As an example, the radiation hardness of high-voltage CMOS pixel sensors has been tested to a fluence of \SI{1.9d15}{n_{eq}\per\cm^{2}} \cite{Benoit_2018} \cite{Anders_2018}. Currently, an extensive irradiation campaign is ongoing, aiming to validate the radiation tolerance of the data transmission chain \cite{Franconi:2773360} for the read-out of the  Inner Tracker (ITk) pixel detector of the ATLAS experiment \cite{Collaboration_2008}. To this end, diverse components such as cable shieldings, connectors and electrical cables have been irradiated. Figure \ref{fig:twinax} (left) shows the setup used to irradiate the cables connecting the detector front-end modules to the opto-electronic converters \cite{Halser:2773815}. To achieve a uniform dose along the cable length, the cable was wrapped on a spool mounted on a motor, so that the spool could rotate during the irradiation. The spool was enclosed in a metallic container coupled directly to the BTL after the extraction window, in order to limit the amount of radioactivity induced in air. The right-hand part of the figure corresponds to the change in transfer function ($\mathrm{S_{21}}$) of the cable due to the irradiation, measured using a Vector Network Analyser. 
The facility has also been used to test the radiation hardness of components for space missions such as the  Neutral gas and Ion Mass spectrometer (NIM) onboard ESA's JUpiter ICy moon Explorer (JUICE) \cite{JUICE_magnets}, \cite{JUICE_resistors}, and to study radiation effects on  glass windows for optical readout of Gas Electron Multiplier (GEM) based detetors, targeting hadrontherapy applications \cite{Maia_Oliveira_2021}.

\section{Conclusions}

The Bern Medical Cyclotron, an accelerator primarily used for production of medical radioisotopes, is in operation as an irradiation facility thanks to a  Beam Transfer Line (BTL) that can transport the 18-MeV proton beam to an adjacent bunker fully dedicated to research activities. Among other projects, the beam line is used for radiation hardness studies of electronics and materials for scientific applications. A dedicated setup for monitoring the fluence during the irradiation was developed, consisting of two independent instruments: a beam profiler, to measure the transversal beam profile, and a collimator, to measure the beam current. This method was validated in two dedicated measurements. In the first, the proton flux calculated from the beam profile and collimator current was compared with a direct measurement of the beam current using a Faraday cup. An agreement within 10\% between both measurements was obtained for a proton flux ranging from  \SI{5e9}{\cm^{-2}\s^{-1}} to \SI{4e11}{\cm^{-2}\s^{-1}}. In the second, a set of radiochromic films were irradiated to different TID levels. The dose calculated from the beam current and profile and the dose determined by the film colour change were found to be compatible within $15\%$. The facility features also a Physics Laboratory equipped with a wide range of electronics equipment, as well as gamma and alpha spectrometers, thus allowing in depth post-irradiation studies to be performed directly on site. The Bern Medical Cyclotron has hosted different irradiation campaigns for high-energy physics experiments, such as ATLAS, CMS and LHCb, and as well as for astrophysics and medical physics studies.

\acknowledgments

We acknowledge contributions from the LHEP engineering and technical staff who helped us with the mechanical and electronics challenges occurred during the design, the development and installation of the irradiation facility. 
We thank the SWAN Isotopen AG maintenance team for the support and the collaboration.
This work was partially funded by the Swiss National Science Foundation (SNSF) (grants 200020\textunderscore204241, 200020\textunderscore188442, 20FL20\textunderscore201478, 20FL20\textunderscore173601, CR23I2\textunderscore156852, 200021\textunderscore175749, CRSII5\textunderscore180352, CR23I2\textunderscore156852).

\bibliographystyle{JHEP}
\bibliography{article.bib}

\providecommand{\href}[2]{#2}\begingroup\raggedright\begin{thebibliography}{10}

\bibitem{furano2013review}
G.~Furano, R.~Jansen and A.~Menicucci, \emph{Review of radiation hard
  electronics activities at European Space Agency},
  \href{https://doi.org/10.1088/1748-0221/8/02/C02007}{\emph{Journal of
  Instrumentation} {\bfseries 8} (2013) C02007}.

\bibitem{poivey2002radiation}
C.~Poivey, \emph{Radiation Hardness Assurance for Space Systems},  in
  \emph{IEEE NSREC Short Course}, pp.~V1--V57, 2002.

\bibitem{label1998emerging}
K.~A. LaBel, A.~H. Johnston, J.~L. Barth, R.~A. Reed and C.~E. Barnes,
  \emph{Emerging Radiation Hardness Assurance (RHA) issues: a NASA approach for
  space flight programs}, {\emph{IEEE Transactions on Nuclear Science}
  {\bfseries 45} (1998) 2727}.

\bibitem{Bruning:782076}
O.~S. Brüning, P.~Collier, P.~Lebrun, S.~Myers, R.~Ostojic, J.~Poole et~al.,
  \emph{{LHC Design Report}}, CERN Yellow Reports: Monographs. CERN, Geneva,
  2004,
  \href{https://doi.org/10.5170/CERN-2004-003-V-1}{10.5170/CERN-2004-003-V-1}.

\bibitem{Aberle:2749422}
O.~Aberle, I.~Béjar~Alonso, O.~Brüning, P.~Fessia, L.~Rossi, L.~Tavian
  et~al., \emph{{High-Luminosity Large Hadron Collider (HL-LHC): Technical
  design report}}, CERN Yellow Reports: Monographs. CERN, Geneva, 2020,
  \href{https://doi.org/10.23731/CYRM-2020-0010}{10.23731/CYRM-2020-0010}.

\bibitem{8116686}
R.~García~Alía, M.~Brugger, F.~Cerutti, S.~Danzeca, A.~Ferrari, S.~Gilardoni
  et~al., \emph{LHC and HL-LHC: Present and Future Radiation Environment in the
  High-Luminosity Collision Points and RHA Implications},
  \href{https://doi.org/10.1109/TNS.2017.2776107}{\emph{IEEE Transactions on
  Nuclear Science} {\bfseries 65} (2018) 448}.

\bibitem{Moll:2018fol}
M.~Moll, \emph{{Displacement damage in silicon detectors for high energy
  physics}}, \href{https://doi.org/10.1109/TNS.2018.2819506}{\emph{IEEE
  Transactions on Nuclear Science} {\bfseries 65} (2018) 1561}.

\bibitem{DIMITRIEVSKA2020162091}
A.~Dimitrievska and A.~Stiller, \emph{RD53A: A large-scale prototype chip for
  the phase II upgrade in the serially powered HL-LHC pixel detectors},
  \href{https://doi.org/https://doi.org/10.1016/j.nima.2019.04.045}{\emph{Nuclear
  Instruments and Methods in Physics Research Section A: Accelerators,
  Spectrometers, Detectors and Associated Equipment} {\bfseries 958} (2020)
  162091}.

\bibitem{6551318}
M.~Esposito, T.~Anaxagoras, O.~Diaz, K.~Wells and N.~M. Allinson,
  \emph{Radiation hardness of a large area CMOS active pixel sensor for
  bio-medical applications},
  \href{https://doi.org/10.1109/NSSMIC.2012.6551318}{\emph{in 2012 IEEE Nuclear
  Science Symposium and Medical Imaging Conference Record (NSS/MIC)} (2012)
  1300}.

\bibitem{https://doi.org/10.1002/pssa.201800383}
I.~A. Zahradnik, M.~T. Pomorski, L.~De~Marzi, D.~Tromson, P.~Barberet,
  N.~Skukan et~al., \emph{scCVD Diamond Membrane based Microdosimeter for
  Hadron Therapy},
  \href{https://doi.org/https://doi.org/10.1002/pssa.201800383}{\emph{physica
  status solidi (a)} {\bfseries 215} (2018) 1800383}.

\bibitem{VUKOLOV2015177}
K.~Vukolov, A.~Borisov, N.~Deryabina and I.~Orlovskiy, \emph{Development of
  ITER diagnostics: Neutronic analysis and radiation hardness},
  \href{https://doi.org/https://doi.org/10.1016/j.fusengdes.2015.06.153}{\emph{Fusion
  Engineering and Design} {\bfseries 96-97} (2015) 177}.

\bibitem{9217477}
J.~M. Rafí, G.~Pellegrini, P.~Godignon, S.~O. Ugobono, G.~Rius, I.~Tsunoda
  et~al., \emph{Electron, Neutron, and Proton Irradiation Effects on SiC
  Radiation Detectors},
  \href{https://doi.org/10.1109/TNS.2020.3029730}{\emph{IEEE Transactions on
  Nuclear Science} {\bfseries 67} (2020) 2481}.

\bibitem{SRIM}
J.~F. {Ziegler}, M.~D. {Ziegler} and J.~P. {Biersack}, \emph{{SRIM - The
  stopping and range of ions in matter (2010)}},
  \href{https://doi.org/10.1016/j.nimb.2010.02.091}{\emph{Nuclear Instruments
  and Methods in Physics Research B} {\bfseries 268} (2010) 1818}.

\bibitem{cyclotron_website}
``Bern medical cyclotron irradiation facility.''
\newblock
  {https://www.lhep.unibe.ch/research/medical\_applications/index\_eng.html}.

\bibitem{Cyclo}
S.~{Braccini}, \emph{{The new Bern PET cyclotron, its research beam line, and
  the development of an innovative beam monitor detector}},
  \href{https://doi.org/10.1063/1.4802308}{\emph{in American Institute of
  Physics Conference Series} {\bfseries 1525} (2013) 144}.

\bibitem{Scampoli:2011zz}
P.~Scampoli, K.~von Bremen, S.~Braccini and A.~Ereditato, \emph{{The New Bern
  Cyclotron Laboratory for Radioisotope Production and Research}}, {\emph{Conf.
  Proc. C} {\bfseries 110904} (2011) 3619}.

\bibitem{molecules25204706}
N.~P. van~der Meulen, R.~Hasler, Z.~Talip, P.~V. Grundler, C.~Favaretto, C.~A.
  Umbricht et~al., \emph{Developments toward the Implementation of 44Sc
  Production at a Medical Cyclotron},
  \href{https://doi.org/10.3390/molecules25204706}{\emph{Molecules} {\bfseries
  25} (2020) }.

\bibitem{braccini:cyclotrons2019-tua02}
S.~Braccini, C.~Belver-Aguilar, T.~Carzaniga, G.~Dellepiane, P.~Haeffner and
  P.~Scampoli, \emph{{Novel Irradiation Methods for Theranostic Radioisotope
  Production With Solid Targets at the Bern Medical Cyclotron}},
  \href{https://doi.org/10.18429/JACoW-Cyclotrons2019-TUA02}{\emph{in Proc.
  Cyclotrons'19} (2020) 127}.

\bibitem{Belver-Aguilar:2020plz}
C.~Belver-Aguilar, S.~Braccini, T.~Carzaniga, A.~Gsponer, P.~Häffner,
  G.~Molinari et~al., \emph{{Development of Novel Non-Destructive 2D and 3D
  Beam Monitoring Detectors at the Bern Medical Cyclotron}},
  \href{https://doi.org/10.18429/JACoW-IBIC2020-TUPP32}{\emph{{in 9th
  International Beam Instrumentation Conference}} (2020) }.

\bibitem{app10228217}
C.~Belver-Aguilar, S.~Braccini, T.~S. Carzaniga, A.~Gsponer, P.~D. Häffner,
  P.~Scampoli et~al., \emph{A Novel Three-Dimensional Non-Destructive
  Beam-Monitoring Detector},
  \href{https://doi.org/10.3390/app10228217}{\emph{Applied Sciences} {\bfseries
  10} (2020) }.

\bibitem{PMID:24982259}
S.~Braccini, A.~Ereditato, K.~Nesteruk, P.~Scampoli and K.~Zihlmann,
  \emph{Study of the radioactivity induced in air by a 15-MeV proton beam},
  \href{https://doi.org/10.1093/rpd/ncu199}{\emph{Radiation protection
  dosimetry} {\bfseries 163} (2015) 269—275}.

\bibitem{instruments3010004}
K.~P. Nesteruk, L.~Ramseyer, T.~S. Carzaniga and S.~Braccini, \emph{Measurement
  of the Beam Energy Distribution of a Medical Cyclotron with a Multi-Leaf
  Faraday Cup},
  \href{https://doi.org/10.3390/instruments3010004}{\emph{Instruments}
  {\bfseries 3} (2019) }.

\bibitem{BernEnergy}
K.~P. Nesteruk, M.~Auger, S.~Braccini, T.~S. Carzaniga, A.~Ereditato and
  P.~Scampoli, \emph{{A system for online beam emittance measurements and
  proton beam characterization}},
  \href{https://arxiv.org/abs/1705.07486}{{\ttfamily 1705.07486}}.

\bibitem{Haffner:2706007}
P.~Häffner, C.~B. Aguilar, S.~Braccini, P.~Scampoli and P.~A. Thonet,
  \emph{{Study of the Extracted Beam Energy as a Function of Operational
  Parameters of a Medical Cyclotron}},
  \href{https://doi.org/10.3390/instruments3040063}{\emph{Instruments}
  {\bfseries 3} (2019) 63}.

\bibitem{Cyclo_LC}
M.~Auger, S.~Braccini, A.~Ereditato, K.~P. Nesteruk and P.~Scampoli, \emph{Low
  current performance of the Bern medical cyclotron down to the {pA} range},
  \href{https://doi.org/10.1088/0957-0233/26/9/094006}{\emph{Measurement
  Science and Technology} {\bfseries 26} (2015) 094006}.

\bibitem{UniBeam}
M.~Auger, S.~Braccini, T.~Carzaniga, A.~Ereditato, K.~Nesteruk and P.~Scampoli,
  \emph{A detector based on silica fibers for ion beam monitoring in a wide
  current range},
  \href{https://doi.org/10.1088/1748-0221/11/03/p03027}{\emph{Journal of
  Instrumentation} {\bfseries 11} (2016) P03027}.

\bibitem{DPACE}
``D-PACE.''
\newblock https://www.d-pace.com/.

\bibitem{NISTdatabase}
``NIST database.''
\newblock https://dx.doi.org/10.18434/T4NC7P.

\bibitem{FEI4_irr}
{\scshape ATLAS} collaboration, \emph{{Irradiation induced effects in the FE-I4
  front-end chip of the ATLAS IBL detector}},
  \href{https://doi.org/10.1109/NSSMIC.2016.8069865}{\emph{{in Proceedings,
  2016 IEEE Nuclear Science Symposium and Medical Imaging Conference: NSS/MIC
  2016: Strasbourg, France}} (2016) 1}
  [\href{https://arxiv.org/abs/1611.00803}{{\ttfamily 1611.00803}}].

\bibitem{FWT}
``Far West Technologies.''
\newblock http://www.fwt.com/racm/fwt60ds.htm.

\bibitem{Benoit_2018}
M.~Benoit, S.~Braccini, G.~Casse, H.~Chen, K.~Chen, F.~D. Bello et~al.,
  \emph{Testbeam results of irradiated ams H18 {HV}-{CMOS} pixel sensor
  prototypes},
  \href{https://doi.org/10.1088/1748-0221/13/02/p02011}{\emph{Journal of
  Instrumentation} {\bfseries 13} (2018) P02011}.

\bibitem{Anders_2018}
J.~Anders, M.~Benoit, S.~Braccini, R.~Casanova, H.~Chen, K.~Chen et~al.,
  \emph{Charge collection characterisation with the Transient Current Technique
  of the ams H35DEMO {CMOS} detector after proton irradiation},
  \href{https://doi.org/10.1088/1748-0221/13/10/p10004}{\emph{Journal of
  Instrumentation} {\bfseries 13} (2018) P10004}.

\bibitem{Franconi:2773360}
{L. Franconi (in press)}, \emph{{The Opto-electrical conversion system for the
  data transmission chain of the ATLAS ITk Pixel detector upgrade for the
  HL-LHC}},  in \emph{Proceedings of International Conference on Technology and
  Instrumentation in Particle Physics 2021}, (Online), Jun, 2021,
  \href{https://cds.cern.ch/record/2773360}{https://cds.cern.ch/record/2773360}.

\bibitem{Collaboration_2008}
{The ATLAS Collaboration}, \emph{The {ATLAS} Experiment at the {CERN} Large
  Hadron Collider},
  \href{https://doi.org/10.1088/1748-0221/3/08/s08003}{\emph{Journal of
  Instrumentation} {\bfseries 3} (2008) S08003}.

\bibitem{Halser:2773815}
{L. Halser (in press)}, \emph{{Irradiation studies at the Bern cyclotron for
  the ATLAS ITk upgrade}},  in \emph{Proceedings of International Conference on
  Technology and Instrumentation in Particle Physics 2021}, (Online), Jun,
  2021,
  \href{https://cds.cern.ch/record/2773815}{https://cds.cern.ch/record/2773815}.

\bibitem{JUICE_magnets}
D.~{Lasi}, S.~{Meyer}, D.~{Piazza}, M.~{Lüthi}, A.~{Nentwig}, M.~{Gruber}
  et~al., \emph{Decisions and Trade-Offs in the Design of a Mass Spectrometer
  for Jupiter's Icy Moons},
  \href{https://doi.org/10.1109/AERO47225.2020.9172784}{\emph{in 2020 IEEE
  Aerospace Conference} (2020) 1}.

\bibitem{JUICE_resistors}
D.~{Lasi}, M.~{Tulej}, M.~B. {Neuland}, P.~{Wurz}, T.~S. {Carzaniga}, K.~P.
  {Nesteruk} et~al., \emph{Testing the Radiation Hardness of Thick-Film
  Resistors for a Time-Of-Flight Mass Spectrometer at Jupiter with 18 MeV
  Protons}, \href{https://doi.org/10.1109/NSREC.2017.8115474}{\emph{in 2017
  IEEE Radiation Effects Data Workshop (REDW)} (2017) 1}.

\bibitem{Maia_Oliveira_2021}
A.~M. Oliveira, S.~Braccini, P.~Casolaro, N.~Heracleous, J.~Leidner, I.~Mateu
  et~al., \emph{Radiation-induced effects in glass windows for optical readout
  {GEM}-based detectors},
  \href{https://doi.org/10.1088/1748-0221/16/07/t07009}{\emph{Journal of
  Instrumentation} {\bfseries 16} (2021) T07009}.

\end{thebibliography}\endgroup

\end{document}